\documentclass[10pt,aps,twocolumn,showpacs,pra]{revtex4}
\usepackage{amssymb,amsfonts,amsmath,wrapfig,mathrsfs,mathbbol}
\usepackage{graphicx}
\begin{document}

\title{Photoionization with Orbital Angular Momentum Beams}

\author{A. Pic\'{o}n$^{1,4}$, J. Mompart$^{1}$, J. R. V{\'a}zquez de Aldana$^{2}$, L. Plaja$^{2}$, G. F. Calvo$^{3}$, and L. Roso$^{2}$}

\address{$^{1}$Grup d'\`{O}ptica, Universitat Aut\`{o}noma de Barcelona, E-08193 Bellaterra (Barcelona), Spain}

\address{$^{2}$Servicio L{\'a}ser, Universidad de Salamanca, E-37008 Salamanca, Spain}

\address{$^{3}$ Departamento de Matem\'aticas, ETSI Industriales \& IMACI-Instituto de Matem\'atica Aplicada a la Ciencia y la Ingenier\'{\i}a, Universidad de Castilla-La Mancha, E-13071 Ciudad Real, Spain}

\address{$^{4}$JILA, University of Colorado, Boulder 80309-0440, USA (actual address)} 

\date{\today}

\begin{abstract}
Intense laser ionization expands EinsteinÕs photoelectric effect rules giving a wealth of phenomena widely studied over the last decades. In all cases, so far, photons were assumed to carry one unit of angular momentum. However it is now clear that photons can possess extra angular momentum, the orbital angular momentum (OAM), related to their spatial profile. We show a complete description of photoionization by OAM photons, including new selection rules involving more than one unit of angular momentum. We explore theoretically the interaction of a single electron atom located at the center of an intense ultraviolet beam bearing OAM, envisaging new scenarios for quantum optics. 
\end{abstract}

\maketitle


During the history of Physics, Light-Matter interaction has been the fundamental path towards understanding new phenomena and testing some essential theories, as in the case of photoionization. Approximately until 1992, all physical theories have described light according to three features: energy, linear momentum and polarization. The latter, which is purely related to the electric field direction of the propagating light, yields an effective light angular momentum. But, as recognized by Allen and coworkers [1], light possess another degree of freedom: the orbital angular momentum (OAM), which rather than being associated with polarization, it is related to the spatial profile of light. This newborn degree of freedom has kindled a huge activity in different lines of research, ranging from micro and nanoparticle trapping~\cite{Padgett,Bhattacharya,Rev_Padgett_2008} to quantum state engineering in Bose-Einstein condensates~\cite{Andersen}, multiphoton entanglement~\cite{Mair,Kozuma,CalvoPRA07} for quantum information applications and molecular spectroscopy~\cite{Nulty}. Recently, both femtosecond and high-power OAM beams have been generated experimentally using holographic plates \cite{Mariyenko_Creation_2005,Sola_High_2008}. In this work, we revisit the Einstein photoionization scenario~\cite{Einstein}, but now taking into account the orbital angular momentum of light. This allows us to unveil new phenomena beyond the standard photoionization. 
\par
Photoionization has attracted a broad interest both from a fundamental theoretical point of view and from the standpoint of applications. Laser photoionization, particularly strong field photoionization, has been a very active research topic over the last decades. Many new effects have been reported, such as ATI (Above Threshold Ionization), tunnel ionization, high-order harmonic generation, etc.~\cite{Blaga,Corkum} Besides, much effort has been devoted in the recent years to the possible inhibition of photoionization at high frequencies and for very strong fields. Experimental activity on ultra-strong field ionization has recently reached the relativistic domain~\cite{Mourou}. All this photoionization literature can be classified into two regimes: the electric-dipole regime (where the magnetic field of the light can be neglected) and the non-dipole regime. In the electric-dipole regime, light beams carry the standard angular momentum (one $\hbar$ unit), and the selection rules avoid the possibility of one-photon excitation of atomic transitions with angular momentum variation larger than one $\hbar$. However, one can overcome this limitation by considering multi-photon effects with very intense lasers. In the non-dipole regime, the light-atom interaction is more complex, exciting not only atomic transitions with angular momentum change equal to one unit $\hbar$, but also atomic transitions with larger angular momentum variation. In this work we present for the first time photoionization with light beams carrying OAM which give rise to new selection rules out of both the electric-dipole and the non-dipole regime, opening new perspectives for atomic transition excitations. 
\par
Much numerical work has been devoted to 3D photoionization studies. Strong field photoionization implies ionized electrons that escape at high energies, thus requiring large and dense numerical networks (to describe both electrons far away from the atomic core and energetic electrons driven by the laser). When a linearly polarized laser light is used, numerical solutions are relatively simple due to the cylindrical symmetry of the problem and only two dimensional numerical grids are needed. However, there exist relevant examples where such cylindrical symmetry is not possible~\cite{Javi}, and a true three dimensional numerical grating is needed. This  demands a much larger complexity of the numerical simulation. Here, we present an accurate description of the atomic photoionization induced by a beam carrying OAM. This scenario requires a true 3D simulation to fully account for both the three-spatial dimensions of the electron quantum state and the 3D-spatial profile of the beam (including the transverse profile related to the OAM). 
\par
In this paper we address the interaction of a pulse beam carrying OAM with the simplest atom: hydrogen. This scenario provides the paradigm for fundamental questions: Is the OAM of light transferred to the electron quantum state? There are still open questions about the angular momentum transfer between matter and light~\cite{Padgett, Barnett, VanEnk,Jauregui, Alexandrescu}. The angular momentum of light can be separated in OAM and spin momentum in the paraxial regime, but how are the OAM and the spin momentum of light transferred to the matter? In some previous works \cite{VanEnk,Jauregui,Alexandrescu}, where an ensemble of atoms was considered, it was shown how the OAM can be coupled to the center of mass of the atomic ensemble. The same interaction description out of the paraxial regime \cite{Jauregui} is more plentiful. In contrast with previous works, we are considering the interaction of a Laguerre-Gaussian intense laser beam with a single quantum state placed near its optical vortex, taking into account the general form of the quantum electromagnetic-interaction-field hamiltonian, without neglecting any magnetic term and without restricting to any multipolar approximation in the transverse plane, going beyond into the comprehension of the OAM light coupling with a quantum system. To clarify the picture, we investigate photoionization with OAM beams in the Schr\"odinger regime, using both analytical and numerical tools.

\begin{figure}[htbp]
\centering\includegraphics[width=7cm]{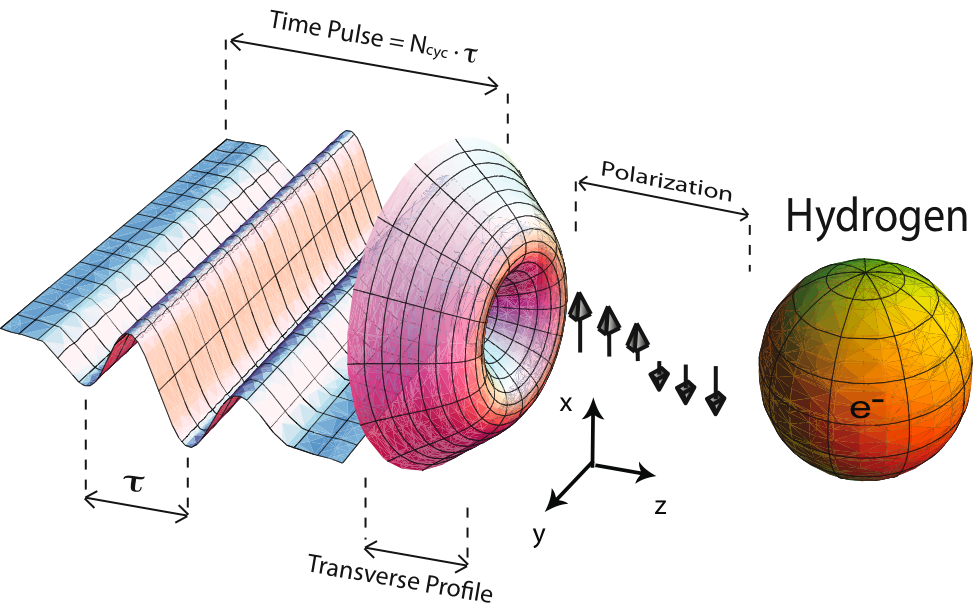}
\caption{\small {\bf Light-Matter Interaction Scheme}. The addressed problem consists of: A temporal pulse with a well-defined polarization and a transverse profile that takes into account the OAM. The initial state of the electron corresponds to the fundamental state of the hydrogen atom.}
\label{fig:Sketch}
\end{figure}


\section{Light-Matter Interaction Scheme}

We begin by considering a pulse beam propagating along the $z$-direction with a temporal envelope wave parameterized by a quadratic sinus (see Fig. \ref{fig:Sketch}). This temporal envelope has a frequency $\omega_{e} = \pi/N_{cyc}\tau$, where $N_{cyc}$ and $\tau$ are the cycle number and the period of the carrier wave, respectively. A hydrogen atom is assumed to be localized at the origin of the reference system and experiences a vector potential, associated to the pulse, of the form 
\begin{eqnarray}\label{VectorPotential}
{\bf A}_{\ell}({\bf r},t) \!&=&\! {\bf A}_{o} \, w_{0} \sin^{2}\left(\frac{\omega_{e}}{c}(z+a_{o} )-\omega_{e}t \right) \nonumber \\ \!  &\times&\!  \Bigl[  \theta(z-ct+\pi c/\omega_{e} + a_{o})-\theta(z-ct+a_{o})\Bigr] \nonumber \\ 
\!&\times&\! \Bigl[\, e^{i\frac{\omega}{c} (z-ct)}LG_{\ell,p}(\rho,\phi,z;\frac{\omega}{c})+ \;\textrm{c. c.}\,\Bigr] \ ,
\label{VectorPotential}
\end{eqnarray}
where $\theta$ is the step function, $a_{o}$ the Bohr radius, $c$ the speed of light,   $\omega$ the carrier wave frequency, and ${\bf A}_{o}$ the amplitude of the wave (it includes the polarization state). The transverse  spatial structure of the pulse beam is accounted by the functions $LG_{\ell,p}(\rho,\phi,z;\omega/c)$; the Laguerre-Gaussian modes~\cite{Allen}. They are characterized by a width $w_{0}$ at $z=0$ (the beam waist), and by the indices $\ell=0,\pm 1,\pm 2,\ldots$ and $p=0, 1, 2,\ldots$, representing the, so-called, winding (or topological charge) and the number of nonaxial radial nodes of the mode, respectively.  Laguerre-Gaussian modes  contain an azimuthal phase $e^{i\ell\phi}$ which gives rise to a discrete OAM of $\ell\hbar$ units per photon along their propagation direction. The complete spatial spectrum of scalar wave fields prepared in arbitrary superpositions of Laguerre-Gaussian (or other paraxial) modes can be measured by using a simple interferometric scheme~\cite{Calvo08}. The fact that the associated electric field amplitude now depends on the transverse position, in contrast with plane waves, will be shown below to give rise to unexpected phenomena. 
\par
We assume that the hydrogen nucleus is unaffected by the electromagnetic field. If so, the electron-field coupling evolution will be described by the following Schr{\"o}dinger equation
\begin{eqnarray}\label{Schrodinger}
i \hbar \frac{\partial}{\partial t} \psi ({\bf r}, t) = \mathcal{\hat{H}} ({\bf r}, {\bf p}, t) \psi ({\bf r}, t) = \nonumber \\ \left[ \frac{1}{2m} \left( \hat{\bf p} - q\,\hat{\bf A} ({\bf r}, t)  \right)^{2} + q\, \hat{V}(r)\right] \psi ({\bf r}, t)\; ,
\end{eqnarray}
where $\psi ({\bf r}, t)$ is the electron quantum state, $\hat{V}(r)$ the Coulomb potential originated by the hydrogen nucleus, $m$ the electron mass, $q$ the electron charge, $\hat{\bf A} ({\bf r}, t)$ the vector potential, which in our case is given by expression~(\ref{VectorPotential}), and $ \hat{\bf p} \equiv -i\hbar{\boldsymbol \nabla}$ the linear momentum operator, satisfying the canonical commutation relation $[\hat{\bf x}, \hat{\bf p}] = i\hbar$. 


\section{Selection Rules with OAM}

We first extract a new set of selection rules for the interaction of atoms with OAM beams and point out the essential differences with pulse beams consisting of plane waves. The Hamiltonian (\ref{Schrodinger}) of the system can be expressed in the usual form; $\mathcal{\hat{H}}=\mathcal{\hat{H}}_{0}+\mathcal{\hat{H}_{I}}+\mathcal{\hat{H}_{II}}$, where $\mathcal{\hat{H}}_{0}$ is the free part, whereas $\mathcal{\hat{H}_{I}}\equiv -q \; (\hat{\bf p}\cdot\hat{\bf A} ({\bf r}, t)+\hat{\bf A} ({\bf r}, t)\cdot\hat{\bf p})/2m$ and $\mathcal{\hat{H}_{II}}\equiv q^{2}\hat{\bf A}^{2} ({\bf r}, t)/2m$ refer to the interaction parts.  Representing the quantum state in a spherical basis $\psi ({\bf r})= \sum_{L,M} u_{L,M} (r) Y^{M}_{L}(\theta,\varphi)$ (all radial dependence is in the functions $u_{L,M} (r)$, while the angular dependence remains in the spherical harmonic functions $Y^{M}_{L}(\theta,\varphi)$ instead), the first interaction contribution can be written as
\begin{eqnarray} \label{1IntH_element}
\langle \psi_{f}\vert\mathcal{\hat{H}_{I}} \vert \psi_{i} \rangle= i\frac{q}{\hbar}(E_{i}-E_{f}) \, \langle \psi_{f}\vert \, {\bf r}\cdot\hat{\bf A} ({\bf r}, t) \vert \psi_{i} \rangle \; .
\end{eqnarray}
Here, $E_{i}$ and $E_{f}$ are the unperturbed energies of the initial and final states, respectively. Taking into account the vector potential given by (\ref{VectorPotential}), and within the dipolar ($\lambda \gg a_{o}$) and the transverse spatial ($w_{0} \gg \lambda$) approximations, we derive the following set of selection rules for beams carrying any arbitrary $\ell$ units of OAM (see the Appendix):
\begin{eqnarray}
\vert \Delta L\vert \leq \vert \ell \vert + 1 , \; \Delta M = \ell \pm 1, \; \text{with} \; \Delta L + \vert \ell \vert + 1 \; \text{even} ,  \label{selection_rules}
\end{eqnarray}
where we have assumed that the quantization axis is along the beam propagation direction. In contrast with plane waves (for which $\ell=0$ and $\vert \Delta L\vert = 1$), significant variations of the angular momentum are to be expected. In terms of photons, the selection rules (\ref{selection_rules}) can be conceived as the absorption of photons carrying a total angular momentum $j=\ell + s$ in the propagation direction, where $s$ indicates the polarization part (or spin momentum, $s=\pm1$ for right- and left-circular polarization). We would like to remark that these selection rules originate from the transverse profile, despite the dipolar approximation. Moreover, the second contribution for the interaction Hamiltonian $\mathcal{\hat{H}_{II}}$ yields, in the case of plane waves, a constant term, producing a ponderomotive force \cite{Hilbert09}. In our case, due to the transverse profile of the beam, this Hamiltonian part produces two contributions. One acting as a ponderomotive force, while the other one, remarkably, gives rise to new selection rules (see the Appendix):
\begin{eqnarray} \label{selection_rules_2}
\vert \Delta L\vert \leq 2\vert \ell \vert  , \; \Delta M = 2\ell , \; \textrm{with} \; \Delta L  \; \textrm{even} .
\end{eqnarray}
We should remark that the domain of applicability of selection rules (\ref{selection_rules}) and (\ref{selection_rules_2}) extends beyond the photoionization problem.

\begin{figure}[htbp]
\centering\includegraphics[width=7cm]{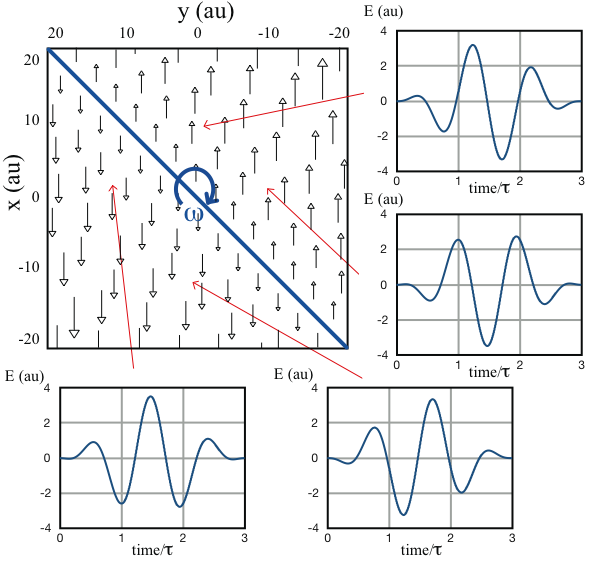}
\caption{\small {\bf Carrier Envelope Phase for a Laguerre-Gaussian Beam}. In the main figure, the polarization of the pulse beam is represented for a transverse plane at $z=0$, just when the pulse is maximum. The atom is centered at the origin (coincident with the beam vortex), where the electric field is zero. The arrow size is proportional to the electric field amplitude strength, which increases linearly in the radial direction up to amplitudes of about 5 au in the boundary distance, 20 au (1nm), and it is harmonically modulated by the azimuthal position. The blue line is the nodal line of the electric field, and rotates with the carrier wave frequency $\omega$, setting the electric field distribution during the pulse interaction. The temporal dependence of the pulse beam is also shown at four different positions of the transverse plane. Notice that the carrier envelope phase (CEP) differs from the azimuthal dependence, not the radial.}
\label{fig:polarization}
\end{figure}


\section{Hydrogen Simulations}

Selection rules (\ref{selection_rules}) and (\ref{selection_rules_2}) constitute our first main result. To gain a more complete picture, we now proceed with the exact description of the hydrogen electron state ionization by beams carrying OAM. Prior to the interaction, we assume the electron to be in the ground state, $\psi_{i} ({\bf r})= 1/\sqrt{\pi a_{o}^{3}}e^{-r/a_{o}}$. We simulate the evolution of the electron quantum state when the incoming pulse has $N_{cyc}=3$, $\ell=1$, $p=0$, an angular frequency $\omega = 1$ au (atomic units where $\hbar=m=q=1$, and $\omega=2\pi \times 6.57\cdot10^{15}$ s$^{-1}$, ultraviolet), a period $\tau = 2\pi$ au ($152$ as), and two possible polarizations: linear (in the $x$-direction) and left-handed.  We choose a beam waist to satisfy the paraxial regime, $w_{0}=9\cdot 10^{4}$ au ($4.79$ $\mu$m), which is much larger than the characteristic size of the atom ($w_{0}\gg a_{o}$). Our atom is centered at the beam vortex, interacting with its vicinity, where the electric field amplitude is much weaker than the maximum one (reached at a distance $w_{0}/\sqrt{2}$). This imposes the need for very intense lasers; in the proximity of the vortex singularity the electric field amplitude increases linearly (when $\vert\ell\vert =1$). For example, an electric field $E\sim A_{o}\, \omega$ of about $10^{4}$ au ($10^{13}$ V/cm), would give rise to 5 au amplitudes (during the pulse peak) at distances of about 20 au (1 nm) from the singularity. In order to clarify the structure of the electric field, in Fig. \ref{fig:polarization} we represent the polarization of the electric field in the transverse plane $z=0$, when the pulse achieves its maximum value. Figure \ref{fig:polarization} also plots the pulse beam with respect to time at four different positions in the transverse plane. Notice the variation of the carrier envelope phase (CEP) depending on the azimuthal position. In fact, all the possible CEPs are encompassed in a circle around the singularity, in contrast with standard few cycle pulses technology~\cite{Brabec}, where much effort has been done to lock the carrier-envelope offset.

\begin{figure}[htbp]
\centering\includegraphics[width=7cm]{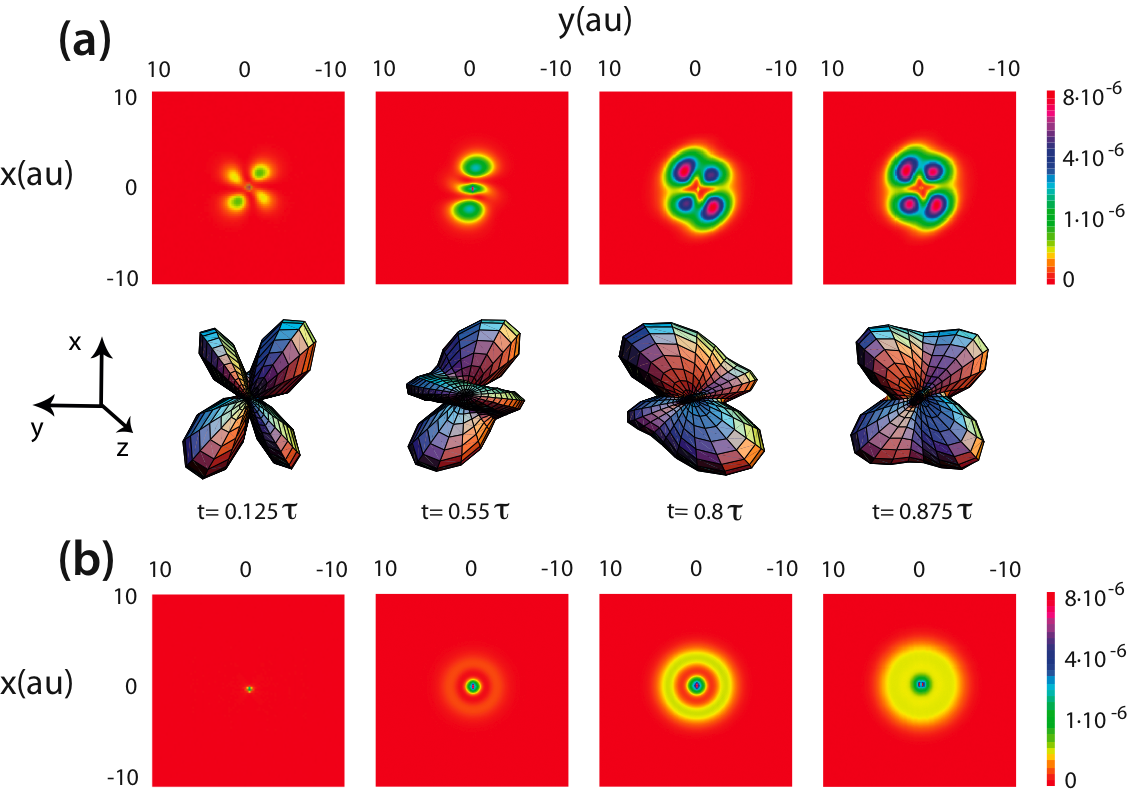}
\caption{\small {\bf Initial Ionization of the Electron Quantum State.} (a) In the upper row, projection of the excited state onto the plane $xy$ ($\int dz \, \vert \delta\psi (x,y,z) \vert^{2}$) at four different times when the pulse beam is linearly polarized. The electron is beginning to be ionized in the first cycle of the pulse beam (with a period of $\tau$=152 as). In the middle row, the superposition of spherical harmonics with widths given by the numerical simulation is represented; they show perfect agreement with the selection rules (\ref{selection_rules},\ref{selection_rules_2}). (b) Same as upper row in (a) but for a left-circularly polarized pulse beam. }
\label{fig:ft_dens}
\end{figure}

Since the electric field has a frequency $\omega=1$ au (larger than the bound hydrogen energy 0.5 au), ionization is to be expected. In fact, after the first pulse cycle we verify that 52\% of the quantum electron state is ionized when the beam is linearly polarized, and 31\% when the beam is circularly polarized. Note that Eq. (\ref{Schrodinger}) yields a total photoionization probability that depends non-linearly on the light polarization. We can express the electron quantum state at each time as $\vert \psi (t)\rangle = \alpha\, \vert \psi_{i}\rangle + \vert \delta\psi (t)\rangle$, where $\delta\psi$ is the excited state part. The excited state function can be decomposed in an unbound spherical basis $\delta \psi ({\bf r})= \sum_{L,M} u_{L,M} (r) Y^{M}_{L}(\theta,\varphi)$. Our initial electron state has full spherical symmetry, belonging to the spherical harmonic $Y^{0}_{0}$. However, after the interaction with the pulse, the electron state excites different spherical harmonics. In Fig. \ref{fig:ft_dens} the projection of the excited state onto the $xy$-plane is depicted during the first cycle of the pulse beam, being a superposition of spherical harmonics obeying the selection rules (\ref{selection_rules}). Also, depending on the input polarization, the electron evolution varies noticeably.

\begin{figure}[htbp]
\centering\includegraphics[width=7cm]{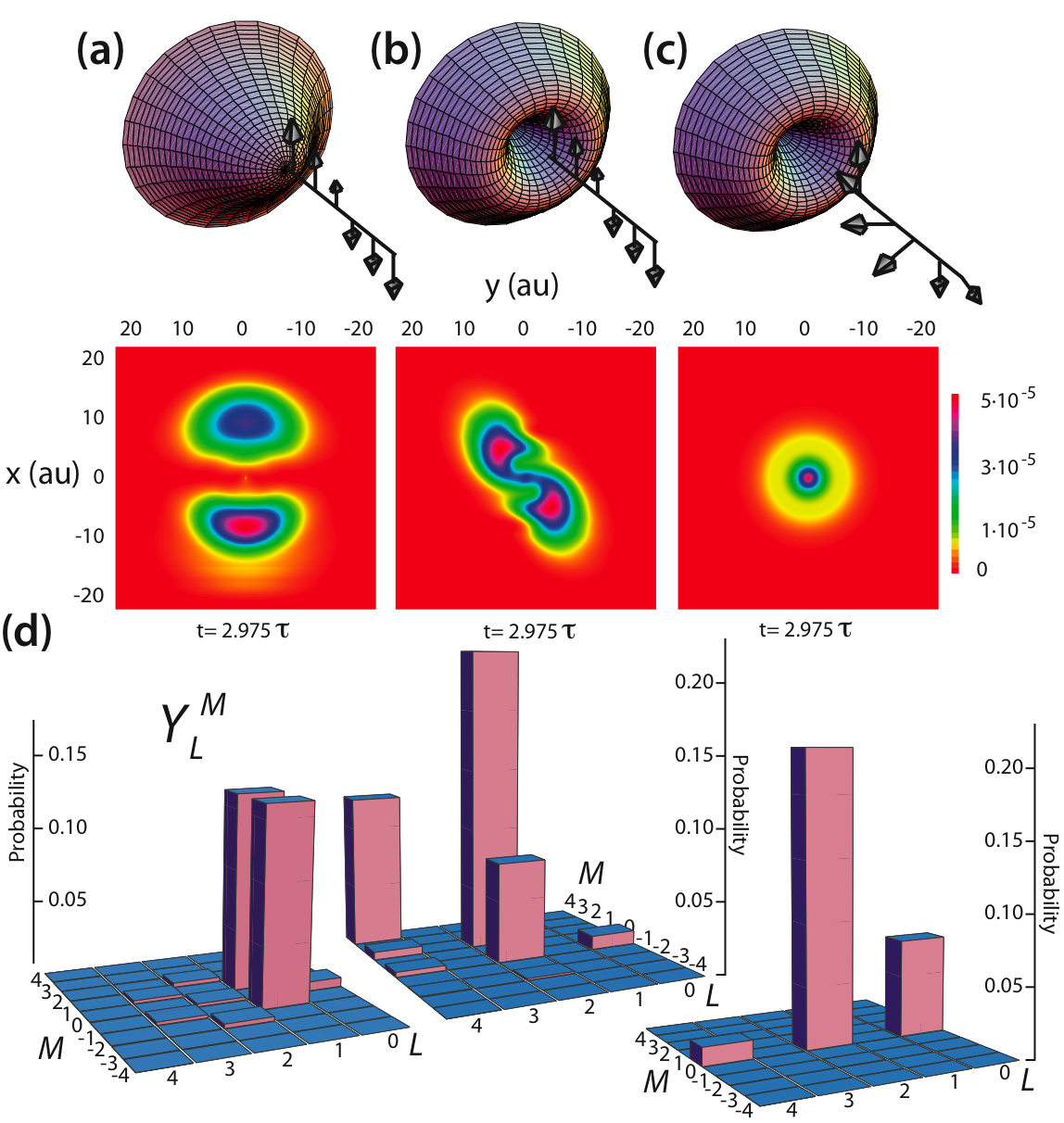}
\caption{\small {\bf Final Quantum Electron State and their Spherical Harmonic Spectrum.} Projection of the excited state ($\int dz \, \vert \delta\psi (x,y,z) \vert^{2}$) onto the plane $xy$ after the interaction with the pulse beam for: (a) a Gaussian mode, (b) a Laguerre-Gaussian mode linearly polarized and (c) a Laguerre-Gaussian mode circularly polarized. (d) Spectrum of spherical harmonics for the three different cases (a), (b) and (c).}
\label{fig:densxz}
\end{figure}

By resorting to numerical approaches, we could accurately evaluate the projections of the excited states onto the spherical harmonics $Y^{M}_{L}$ and extract the corresponding probabilities $P_{L,M} = \int dr \; r^{2} \vert u_{L,M}(r) \vert^{2}$. Using this numerical method, the widths of the spherical harmonic superpositions at  different times have been derived, showing excellent agreement with the electron state evolution, as represented in Fig. \ref{fig:ft_dens}(a). Moreover, we have analyzed the lowest spherical harmonic content of the final excited electron state (see Fig. \ref{fig:densxz}) in three scenarios: (i) with a Gaussian pulse beam (in the transverse spatial approximation, it could be considered as a plane wave) linearly polarized in the $x$-direction (the electron state is ionized about 30\%);  (ii) the case for a pulse spatially modulated by a Laguerre-Gaussian mode, linearly polarized in the $x$-direction;  (iii) when the Laguerre-Gaussian pulse is left-circularly polarized. The main remark is that the spherical harmonics $Y^{1}_{1}$ and $Y^{-1}_{1}$ are most efficiently excited by the plane-wave-like pulse, in striking contrast with the Laguerre-Gaussian scenario, where no such excitation exists. If the Laguerre-Gaussian pulse is linearly polarized, then, $Y^{0}_{2}$, $Y^{2}_{2}$ and $Y^{4}_{4}$ are the most occupied states, whereas if it is circularly polarized, $Y^{0}_{0}$ and $Y^{0}_{2}$ are the most relevant. There is a small contribution from $Y^{0}_{4}$, as the electron is less ionized. We emphasize that these results are in accordance with the derived selection rules (\ref{selection_rules}) and (\ref{selection_rules_2}).

\begin{figure}
\centering\includegraphics[width=7cm]{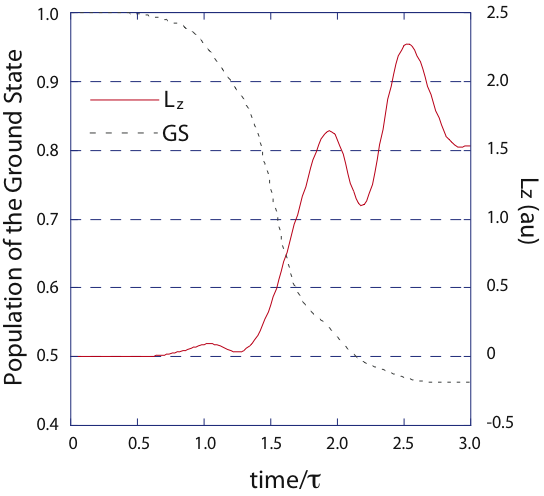}
\caption{\small {\bf Temporal Evolution of the Electron Angular Momentum}. Solid red curve represents the OAM of the electron in the $z$-direction during interaction, parameters as in Fig. \ref{fig:ft_dens}(a). Notice that the electron begins with null OAM and after the pulse beam it gains up to 1.53 au (1.53$\,\hbar$). Dash gray curve represents the population of the ground state. As time increases, it is ionized up to 0.52 of the initial population.}
\label{fig:Lx}
\end{figure}

It is interesting to examine the total angular momentum transferred to the quantum electron state. First, we calculate the mean value of the electron OAM, $\langle \hat{\bf L} \rangle =  \langle \hat{\bf r} \times (\hat{\bf p} - q\,\hat{\bf A})  \rangle$, during its evolution. Figure \ref{fig:Lx} shows the time evolution of the orbital angular momentum along the $z$-direction in the same case of Fig. \ref{fig:ft_dens}(a), while a depopulation of the fundamental state occurs. The electron starts in the ground state, with zero OAM. As the pulse begins to interact with the electron, the OAM of the latter in the $z$-direction oscillates, but notice that at the end of the pulse, the electron quantum state gains a finite amount of OAM: 1.53 au ($1.53\,\hbar$). There is no OAM contribution in other directions. We expect, as the ground state has no OAM in the absence of a field, that the electron excited states belong to unbound states bearing OAM. On the other hand, when the pulse is left-circularly polarized, as the case of Fig. \ref{fig:ft_dens}(b), the OAM in the $z$-direction is negligible, as it is expected. Regarding the excited state position components, mean values are zero except for the $z$-component, where a small shift of $10^{-2}$ au is present, caused by a non-vanishing magnetic field at the origin.


\section{Discussion}
We present the first work addressing the photoionization process induced by beams carrying OAM. We have found novel selection rules (\ref{selection_rules}) and (\ref{selection_rules_2}) for a pulse beam characterized by a topological charge $\ell$. In addition, other interesting effects have been revealed. If the electron excited state after interaction is let to evolve, it goes away from the origin. In contrast, by introducing a pulse beam with $N_{cyc}=14$, we notice that the excited electron state remains confined, within a radius of 10 au, due to the ponderomotive force induced by the Laguerre-Gaussian profile. We also simulate the case where the atom is displaced 2 au from the origin in the $x$ and $y$-directions, with $N_{cyc}=14$. In both cases, the atom is ionized much faster since the electric field is more intense now, and the excited electron state remains trapped. From evaluations of the mean values of the position component, we have observed an electron motion around the vortex. Varying the polarization of the beam, in particular for left-circular polarization, an ionized state with a ring structure ($\Delta M =0$) is predicted, see Figs. \ref{fig:ft_dens}(b) an \ref{fig:densxz}(c). Furthermore, modifying the initial phase, different ionized state structures can be generated. Thus, by tunning the phase and the polarization, one expects a manipulation of the ionized state. There still remain many open questions, such as the feasibility to achieve high-harmonic generation~\cite{Courtial,Patchkovskii} with OAM, the experimental challenges to extend this phenomena to more complex systems (such as Rydberg atoms) and the possibility to use OAM for nuclear quantum optics applications \cite{Ledingham,Keitel}. For instance, by exploiting the transverse profile, one could address the M1 transition at 3.5 eV of Th-229 \cite{Peik_Nuclear_2003}.


\section{Appendix}

We derive in this section the selection rules for the interaction of a light beam possessing orbital angular momentum with matter. From now on, it is convenient to take the quantization axis in the beam propagation direction, the position vector can thus be written as ${\bf r} = (x, y, z) = r \, (\sin\theta \cos\phi,\sin\theta \sin\phi, \cos\theta)$ and $\rho = r \sin\theta$. Hence, the interaction element given by equation (\ref{1IntH_element}) is proportional to
{\small
\begin{eqnarray*} \nonumber
\langle \Psi_{f}\vert \mathcal{\hat{H}_{I}} \vert \Psi_{i} \rangle \propto \hspace{6cm} \\  \nonumber
A_{o} \int_{0}^{\infty} dr \; r^{3} \; \left( \frac{\sqrt{2}r}{w_{0}}\right)^{\!\vert \ell\vert}\! u^{*}_{L_{f},M_{f}} (r) u_{L_{i},M_{i}} (r) \times \\  \nonumber \left\{ \int d\Omega  \; Y_{L_{f}}^{*M_{f}}\! \left[\alpha \; (\sin\theta)^{\vert \ell \vert+1} \cos\phi \; e^{-i(\omega t - \ell\phi-\chi)}\!\! 
+ \! \textrm{c. c.}  \right]\! Y_{L_{i}}^{M_{i}} \; + \right.\\ \left. \int d\Omega  \; Y_{L_{f}}^{*M_{f}}\! \left[\beta \; (\sin\theta)^{\vert \ell \vert+1} \sin\phi \; e^{-i(\omega t - \ell\phi-\chi)}\! +  \textrm{c. c.}  \right]\! Y_{L_{i}}^{M_{i}} \right\}
 \; . 
\end{eqnarray*}
}
\begin{eqnarray}\, \label{1IntH_element_1}
\end{eqnarray}
Equation (\ref{1IntH_element_1}) can be divided into two parts; one depends exclusively on the radial contribution and the other one on the angular contribution. If we expand the angular contribution of Eq. (\ref{1IntH_element_1}), 
\begin{eqnarray} \nonumber
(\sin\theta)^{\vert \ell \vert+1}\cos\phi \; e^{-i(\omega t - \ell\phi-\chi)}  = \frac{e^{-i(\omega t-\chi)}}{2} \times \\ \nonumber 
\left\{ (\sin\theta)^{\vert \ell \vert+1}e^{i(\ell+1)\phi}+(\sin\theta)^{\vert \ell \vert+1}e^{i(\ell-1)\phi} \right\} \; , \\ \nonumber
(\sin\theta)^{\vert \ell \vert+1}\sin\phi \; e^{-i(\omega t - \ell\phi-\chi)}  = \frac{e^{-i(\omega t-\chi)}}{2i} \times \\ \left\{ (\sin\theta)^{\vert \ell \vert+1}e^{i(\ell+1)\phi}-(\sin\theta)^{\vert \ell \vert+1}e^{i(\ell-1)\phi} \right\} \; ,  \label{angular_terms_1_1}
\end{eqnarray}
we can notice basically two terms that are repeated $(\sin\theta)^{\vert \ell \vert+1}e^{i(\ell+1)\phi}$ and $(\sin\theta)^{\vert \ell \vert+1}e^{i(\ell-1)\phi}$. These two terms can be decomposed in spherical harmonics functions, getting then the selection rules. In order to obtain the spherical harmonic decomposition, we must consider two cases, when $\ell \geq 0$ and $\ell \leq 0$. Let us begin with case $\ell \geq 0$, where the angular terms read as
 \begin{eqnarray}  \label{angular_terms_2_1}
(\sin\theta)^{\vert \ell \vert+1}e^{i(\ell+1)\phi} = (-1)^{\ell} 2^{(\ell+1)}(\ell+1) ! \times \hspace{1.7cm} \nonumber \\ \sqrt{\frac{4\pi}{(2\ell+3)!}} \, Y_{\ell+1}^{\ell+1} \; , \hspace{0.5cm} \\
(\sin\theta)^{\vert \ell \vert+1}e^{i(\ell-1)\phi} = (-1)^{\ell} 2^{(\ell-1)}(\ell-1) ! \sqrt{\frac{4\pi}{(2\ell-1)!}} \times \nonumber \\ \frac{4 \sqrt{5}}{3} \; Y_{\ell-1}^{\ell-1} (Y_{0}^{0}-\sqrt{\frac{1}{5}}Y_{2}^{0}) \; , \hspace{0.5cm}  \label{angular_terms_2_2}
\end{eqnarray}
 where we have used $\sin^{2} \theta = 4 \sqrt{5}(Y_{0}^{0}-Y_{2}^{0}/\sqrt{5})/3$ and $Y_{\ell}^{\ell}=(-1)^{\ell} \sqrt{(2\ell+1)!} \sin^{\ell}\theta \, e^{i\ell\phi} / 2^{\ell}\ell! \sqrt{4\pi}$. The first term (\ref{angular_terms_2_1}), in integral (\ref{1IntH_element_1}), gives rise to the straightforward selection rule: $\vert \Delta L \vert \leq \ell + 1$, $\Delta L +  \ell + 1$ is even and $\Delta M = \ell + 1$. 
In equation (\ref{angular_terms_2_2}) there are two contributions, the first one yields also straightforwardly the selection rule: $\vert \Delta L \vert \leq \ell - 1$, $\Delta L +  \ell - 1$ is even and $\Delta M = \ell - 1$, but the second one is a bit subtle. First of all, we need to decompose the product $Y_{\ell-1}^{\ell-1}Y_{2}^{0}$ into spherical harmonics. Using the following formula;
\begin{eqnarray} \nonumber
Y_{l_{1}}^{m_{1}} \; Y_{l_{2}}^{m_{2}} = \sum_{l=\vert l_{1}-l_{2}\vert}^{l_{1}+l_{2}} \sum_{m=-l}^{m=l} \sqrt{\frac{(2l_{1}+1)(2l_{2}+1)}{4\pi (2l+1)}} \times \\ \langle l_{1},l_{2};0,0\vert l,0\rangle \langle l_{1},l_{2};m_{1},m_{2}\vert l,m\rangle \; Y_{l}^{m} \; , \label{formula_2spherical_expansion}
\end{eqnarray}
where $\langle l_{1},l_{2};m_{1},m_{2}\vert l_{3},m_{3}\rangle$ are the corresponding Clebsch-Gordan coefficients, we know that it is possible to write $Y_{\ell-1}^{\ell-1}Y_{2}^{0} = a\, Y_{\ell+1}^{\ell-1}+b\, Y_{\ell}^{\ell-1}+c\, Y_{\ell-1}^{\ell-1}+d\, Y_{\ell-2}^{\ell-1}+e\, Y_{\ell-3}^{\ell-1}$. We can take $d=e=0$, as $M\leq L$ must be satisfied in a spherical harmonic function. And $b=0$ always, due to $\langle \ell-1,2;0,0\vert \ell,0\rangle=0$. Concretely, 
\begin{eqnarray*}
a= \sqrt{\frac{(2\ell -1)5}{4\pi(2\ell+3)}}  \langle \ell-1,2;0,0\vert \ell+1,0\rangle \times \hspace{1cm} \\ \langle \ell-1,2;\ell-1,0\vert \ell+1,\ell-1\rangle \; , \\
c= \sqrt{\frac{(2\ell -1)5}{4\pi(2\ell-1)}} \langle \ell-1,2;0,0\vert \ell-1,0\rangle \times \hspace{1cm} \\ \langle \ell-1,2;\ell-1,0\vert \ell-1,\ell-1\rangle \; .
\end{eqnarray*}
Hence, the second contribution from Eq. (\ref{angular_terms_2_2}) to the selection rules can be summarized as:  $\vert \Delta L \vert \leq \ell + 1$, $\Delta L +  \ell + 1$ is even and $\Delta M = \ell - 1$.
\par
Now, let us consider the case when $\ell \leq 0$. Proceeding in an analogous way than before, we can now write expressions (\ref{angular_terms_2_1}) and (\ref{angular_terms_2_2}) as:
  \begin{eqnarray}  \label{angular_terms_3_1}
(\sin\theta)^{\vert \ell \vert+1}e^{i(\ell+1)\phi} =  2^{(\vert\ell\vert-1)}(\vert\ell\vert-1) ! \sqrt{\frac{4\pi}{(2\vert\ell\vert-1)!}} \times \nonumber \\ \frac{4 \sqrt{5}}{3} \; Y_{\vert\ell\vert-1}^{\ell+1} (Y_{0}^{0}-\sqrt{\frac{1}{5}}Y_{2}^{0}) \; , \hspace{0.5cm} \\
(\sin\theta)^{\vert \ell \vert+1}e^{i(\ell-1)\phi} =  2^{(\vert\ell\vert+1)}(\vert\ell\vert+1) ! \times \hspace{1.7cm} \nonumber \\ \sqrt{\frac{4\pi}{(2\vert\ell\vert+3)!}} \, Y_{\vert\ell\vert+1}^{\ell-1} \; , \hspace{0.5cm}  \label{angular_terms_3_2}
\end{eqnarray}
where we have used $\sin^{2} \theta = 4 \sqrt{5}(Y_{0}^{0}-Y_{2}^{0}/\sqrt{5})/3$ and $Y_{\ell}^{\ell}=\sqrt{(2\vert\ell\vert+1)!} \sin^{\vert\ell\vert}\theta \, e^{i\ell\phi} / 2^{\vert\ell\vert}\vert\ell\vert! \sqrt{4\pi}$. The second term (\ref{angular_terms_3_2}), in integral (\ref{1IntH_element_1}), yields: $\vert \Delta L \vert \leq \vert\ell\vert + 1$, $\Delta L +  \vert\ell\vert + 1$ is even and $\Delta M = \ell - 1$. 
On the other hand, in Eq. (\ref{angular_terms_3_1}) there are two contributions, the first one giving rise to: $\vert \Delta L \vert \leq \vert\ell\vert - 1$, $\Delta L +  \vert\ell\vert - 1$ is even and $\Delta M = \ell + 1$. In the second one, the product $Y_{\vert\ell\vert-1}^{\ell-1}Y_{2}^{0}$ must be decomposed into spherical harmonics using formula (\ref{formula_2spherical_expansion}). As before, we can write $Y_{\vert\ell\vert-1}^{\ell+1}Y_{2}^{0} = a'\, Y_{\vert\ell\vert+1}^{\ell+1}+b'\, Y_{\vert\ell\vert}^{\ell+1}+c'\, Y_{\vert\ell\vert-1}^{\ell+1}+d'\, Y_{\vert\ell\vert-2}^{\ell+1}+e'\, Y_{\vert\ell\vert-3}^{\ell+1}$, where $d'=e'=0$ owing to $\vert M\vert \leq L$, which must be satisfied in any spherical harmonic function. And $b'=0$ always, due to $\langle \vert\ell\vert-1,2;0,0\vert \vert\ell\vert,0\rangle=0$. Therefore, 
\begin{eqnarray*}
a'= \sqrt{\frac{(2\vert \ell \vert -1)5}{4\pi(2\vert\ell\vert+3)}} \langle \vert\ell\vert-1,2;0,0\vert \vert\ell\vert+1,0\rangle \times \hspace{1cm} \\  \langle \vert\ell\vert-1,2;\ell+1,0\vert \vert\ell\vert+1,\ell+1\rangle \; , \\
c'= \sqrt{\frac{(2\vert \ell \vert -1)5}{4\pi(2\vert\ell\vert-1)}} \langle \vert\ell\vert-1,2;0,0\vert \vert\ell\vert-1,0\rangle \times \hspace{1cm} \\  \langle \vert\ell\vert-1,2;\ell+1,0\vert \vert\ell\vert-1,\ell+1\rangle \; .
\end{eqnarray*}
implying that:  $\vert \Delta L \vert \leq \vert\ell\vert + 1$, $\Delta L +  \vert\ell\vert + 1$ is even and $\Delta M = \ell + 1$.
\par
Generalizing all the last calculations, the selection rules derived from the matrix element (\ref{1IntH_element_1}) for beams carrying any unit of orbital angular momentum are summarized in Eq. (\ref{selection_rules}).
\par
At variance with the plane waves $\ell=0$ ($\vert \Delta L\vert = 1$), we can expect larger exchange of angular momentum. Of course, playing with the beam polarization ($\alpha$ and $\beta$), as we can note in equation (\ref{selection_rules}), we can modify the selection rules. For example, for a right-circular polarization ($\alpha=1$ and $\beta = i$), the only surviving terms are given by expressions (\ref{angular_terms_2_1}) and (\ref{angular_terms_3_1}), restricting $\Delta M = \ell +1$. Analogously, for a left-circular polarization ($\alpha=1$ and $\beta = -i$) the only surviving terms are given by expressions (\ref{angular_terms_2_2}) and (\ref{angular_terms_3_2}), restricting $\Delta M = \ell -1$. Selection rules (\ref{selection_rules}), in photon terms, can be thought as the absorption of a photon carrying a total angular momentum in the propagation direction $m=\ell + s$, where $s$ indicates the polarization part (spin momentum, for right-circular polarization $s=1$ and for left-circular polarization $s=-1$). We would like to remark that these selection rules are exclusive to the transverse profile. Moreover, the second interaction hamiltonian $\mathcal{\hat{H}_{II}}$, in the case of plane waves, is just a constant term, yielding a ponderomotive force. In our case, it is quite different. Analogously to equation (\ref{1IntH_element}), we can write 
\begin{eqnarray} \label{2Interaction_Hamiltonian_3}
\langle \Psi_{f}\vert \mathcal{\hat{H}_{II}} \vert \Psi_{i} \rangle= \frac{q^{2}}{2m} \langle \Psi_{f}\vert  {\bf A}^{2}({\bf r}, t) \vert \Psi_{i} \rangle \; ,
\end{eqnarray}
in order to extract the selection rules. Again, considering the vector potential (1), in which the dipolar and transverse spatial approximation has been taken into account, we can write
\begin{eqnarray} \nonumber
\langle \Psi_{f} \vert \mathcal{\hat{H}_{II}}\vert \Psi_{i} \rangle \propto \hspace{5cm}\\ 
\left(\frac{2}{w_{0}^{2}}\right)^{\!\vert \ell\vert}\! \langle \Psi_{f}\vert  \Bigl[{\bf A}_{o}\, \rho^{\vert l\vert}\; e^{-i(\omega t - \ell\phi-\chi)} +  \textrm{c. c.} \Bigr]^{2}  \vert \Psi_{i} \rangle \; . \label{2Interaction_Hamiltonian_4}
\end{eqnarray}
Eq. (\ref{2Interaction_Hamiltonian_4}), if we expand the quadratic potential, two distinguishable terms appear:
\begin{eqnarray*}
\Bigl[{\bf A}_{o}\, \rho^{\vert l\vert}\; e^{-i(\omega t - \ell\phi-\chi)} +  \textrm{c. c.} \Bigr]^{2} =\hspace{3cm} \\ 
A_{o}^{2} \; r^{2\vert l\vert} \Bigl[ (\alpha^{2}+\beta^{2})(\sin\theta)^{2\vert \ell \vert}\; e^{-i2(\omega t - \ell\phi-\chi)}+\textrm{c. c.}\Bigr]  \\ 
+  (A_{o}\, \rho^{\vert l\vert})^{2}\; ,
\end{eqnarray*}
where the second term does not depend on time, it acts as a well-potential. On the other hand, the first term gives rise to new selection rules as
 \begin{eqnarray*}  \label{angular_terms}
(\sin\theta)^{2\vert \ell \vert}e^{i2\ell\varphi} =  2^{2\ell}(2\ell) ! \sqrt{\frac{4\pi}{(4\ell+1)!}} \, Y_{2\ell}^{2\ell} \; , 
\end{eqnarray*}
yielding the selection rules of Eq. (\ref{selection_rules_2}).


\section{Acknowledgments}
We acknowledge support by the Spanish Ministry of Education and Science under contracts FIS2005-01369, FIS2008-02425, FIS2006-04151, FIS2007-29091-E, and Consolider projects SAUUL and QOIT, CSD2007-00013, CSD2006-00019, and the Catalan, Junta de Castilla y Le{\'o}n, and Junta de Castilla-La Mancha Governments under contracts SGR2005-00358, SA146A08, and PCI08-0093. We also acknowledge R. Corbal{\'a}n for fruitful discussions.



\end{document}